\documentclass[11pt,twoside]{article}
\usepackage{asp2004}
\usepackage{psfig}
\usepackage{epsf}
\usepackage{graphics}
\usepackage{lscape}
\def\edcomment#1{\iffalse\marginpar{\raggedright\sl#1\/}\else\relax\fi}
\reversemarginpar

\begin{document}
\title{A Jet Model of the Galactic Center Nonthermal Radio Filaments}
\author{F. Yusef-Zadeh}
\affil{Dept Physics and Astronomy, Northwestern University, Evanston, IL
60208}
\author{A. K\"onigl}
\affil{Department of Astronomy and Astrophysics, University of
Chicago, 5640 S. Ellis Ave., Chicago, IL 60637}

\begin{abstract} 
Protostellar sources in star
forming regions are responsible for driving jets with flow
velocities ranging between 300 and 400 km s$^{-1}$. This class of jets
consists
of highly collimated outflows which include  thermal knots with number
densities estimated to be
greater than that of their ambient medium.
On the other hand,
extragalactic FR I jets
consist of light fluid with low  Mach number burrowing  through a denser
medium as the magnetized  jets   radiate  nonthermal emission.
Both protostellar and
extragalactic
jets are believed to be launched
by accretion disks.
Here we consider a jet model
in which
the characteristics common to both protostellar and extragalactic jets
are used to explain the origin of  nonthermal filaments in the Galactic
center
region. We argue that these filaments are analogous to FR I extragalactic
sources
but are launched by embedded young stars or clusters of stars in 
star-forming regions.
\end{abstract}
\thispagestyle{plain}

\section{Introduction}

It has been 20 years since the discovery of the nonthermal radio filaments
(NRFs) associated with the Galactic center Arc was first reported
(Yusef-Zadeh, Morris, \& Chance 1984).  These observations showed evidence
of linear, magnetized features running perpendicular to the Galactic plane.  
A number of NRFs with similar characteristics to the prototype NRFs have been
discovered in the intervening years (Yusef-Zadeh 2003 and references
therein). Several models have suggested that the filaments trace the
illuminated component of a large-scale poloidal magnetic field pervasive
throughout the Galactic center region. However, the presence of a number of
filaments oriented at large angles to the normal to the Galactic plane does
not support the above interpretation and indicates a different origin. Unlike
most models that predict a global, static geometry of the magnetic field
around the Galactic center, the model described here argues for a local
origin. In the proposed picture, the NRFs originate in star-forming
regions. The filaments behave like jets extracting mass and energy from
embedded young stars or clusters of stars as the jets propagate in a dense
ISM of the Galactic center region. A more detailed account of this model
will be given elsewhere.

\section{The Jet Hypothesis}

Recent observations of a number of radio filaments found in the Galactic
center region show a wide range of morphological structures (Nord et al.
2003; Yusef-Zadeh, Hewitt, \& Cotton 2004).  The new images show structures
that suggest NRFs are interacting with their surrounding medium as described
here briefly: \noindent (i) The distortion of the filaments from a straight
geometry; (ii) sub-filamentation at the point where the filaments are most
distorted; (iii) an increase in the surface brightness in the middle of the
filaments; (iv)  a gentle curvature and a widening of the filaments;  (v)  
gaps along the length of the filaments.

The interaction hypothesis implies the flow of high velocity nonthermal
material along the filaments. This interpretation is in contrast to earlier
models in which the strong preexisting organized magnetic field lines are
illuminated by relativistic particles propagating at the Alfv\'en speed.
Many of the morphological characteristics described above have also been
observed in FR I sources. The presumably supersonic and super-Alfv\'enic
jets of FR I sources are known to propagate and interact with a denser
material of their ambient medium (Ferrari 1998).  For example, the
decollimation of some of the filaments and their gentle curvature are
analogous to the morphology of bent jets, which has been
interpreted in terms of collisions with clouds. The widening of
one of the isolated NRFs is shown
in Figure 1 [right panel]  as the filament N2 runs almost perpendicular to
the
Galactic plane;
this widening could be explained either by the negative pressure gradient
expected in the direction away from the Galactic plane or by
entrainment of ambient material into the jet. Another example is the lack
of bright shocked emission from a hot spot at the terminus of the filament;
this can be interpreted in terms of the deceleration of light jets by the
surrounding medium, in similar fashion to FR I sources. Yet another example
is the brightness of many NRFs peaking in the middle of the filaments. In
the context of the jet model, the location of the enhanced surface
brightness is where the jet changes its course as a result of an  encounter
with an obstacle; this in turn may lead to shocks, particle
acceleration, and a change in the spectral and magnetic
properties (Clarke, Burns, \& Feigelson 1986). 
Lastly, the bundles of filaments that make
up the Arc and Sgr C are examples in which the filaments broaden and appear
more diffuse with a steeper spectral index in the direction away from the
Galactic plane (Tsuboi et al. 1986). In analogy with FR I jets,  we believe
that these so-called lobes (see
Fig. 1, left panel) are  produced when jets with a low
internal-to-ambient density ratio get decelerated and disrupted
as they propagate away from the Galactic plane (see Fig. 9a of Ferrari 1998).

\begin{figure}
\plotfiddle{zadeh_fig1.ps}{8.0in}{0}{70}{70}{-130}{-30}
\clearpage
\caption{A 20cm continuum image of the northern extension
of the Galactic center Arc [left panel] and  a close-up grayscale view of
N2
filament with the corresponding contours [right panel] (Yusef-Zadeh,
Hewitt and Cotton
2004).}
\plotfiddle{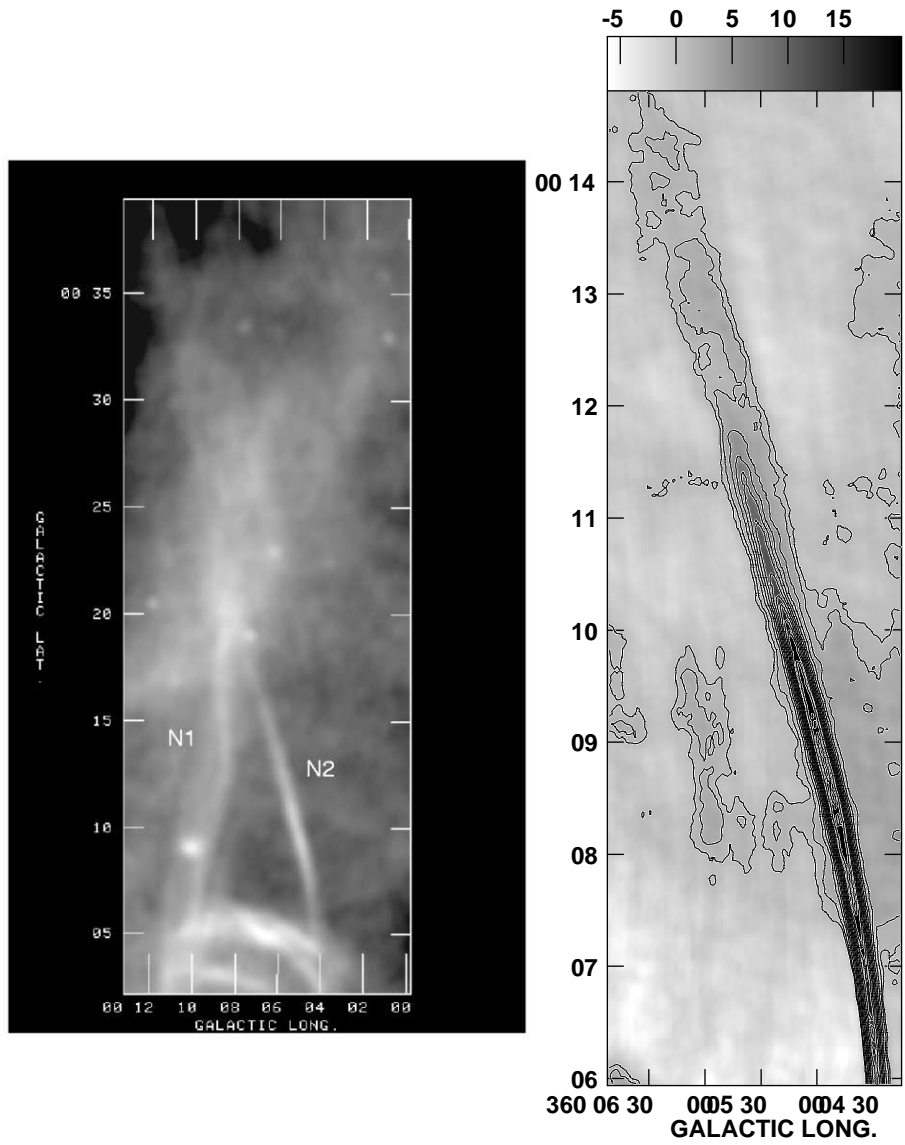}{8.0in}{0}{80}{80}{-190}{700} 
\end{figure}

\end{document}